\documentclass[preprint,twocolumn ]{aastex631}
\usepackage{booktabs}
\usepackage{amsmath}

\begin{document}

\title{Turbulence Properties of Corotating Interaction Regions at 1 AU: a Comprehensive Superposed Epoch Analysis}

\author[0009-0009-0162-2067]{Yuliang Ding}
\affiliation{Department of Earth, Planetary, and Space Sciences, University of California, Los Angeles, Los Angeles, CA 90095, USA}

\author[0000-0002-2582-7085]{Chen Shi}
\affiliation{Department of Earth, Planetary, and Space Sciences, University of California, Los Angeles, Los Angeles, CA 90095, USA}
\affiliation{Department of Physics, Auburn University, Auburn, AL 36849, USA}

\author[0000-0002-2381-3106]{Marco Velli}
\affiliation{Department of Earth, Planetary, and Space Sciences, University of California, Los Angeles, Los Angeles, CA 90095, USA}

\author[0000-0001-9570-5975]{Zesen Huang}
\affiliation{Department of Earth, Planetary, and Space Sciences, University of California, Los Angeles, Los Angeles, CA 90095, USA}

\author[0000-0002-1128-9685]{Nikos Sioulas}
\affiliation{Imperial College London, South Kensington Campus, London SW7 2AZ, UK}



\begin{abstract}
We perform a comprehensive superposed epoch analysis (SEA) of more than 200 corotating interaction regions (CIRs) using WIND spacecraft observations at 1 AU. The stream interfaces are identified by Minimum Variance Analysis (MVA), and turbulence properties are evaluated using wavelet transforms over a wide range of temporal scales. The analysis of normalized cross helicity ($\sigma_c$) and normalized residual energy ($\sigma_r$) reveals distinct turbulence behaviors across frequencies. 
The spectral indices of both magnetic and velocity fluctuations smoothly transition from steeper in the slow wind to shallower in the fast wind, while a localized steepening of the velocity spectra near the interface indicates enhanced dissipation due to compression. Across broad frequency bands, $\sigma_c$ shows a clear dip at the stream interface—signifying increased inward Alfv\'en wave energy—whereas $\sigma_r$ displays a peak–valley–peak structure mainly driven by large-scale velocity shear. In lower-frequency ranges, velocity shear artificially enhances velocity fluctuation energy, producing strong peaks in $\sigma_r$, while higher-frequency ranges show a smooth increase of $\sigma_r$ from slow wind to fast wind. Nearly half of the analyzed CIRs are accompanied by a heliospheric current sheet (HCS), with many HCSs closely aligned with the stream interface, suggesting an intrinsic link between the two structures. 
Our findings offer valuable clues for reconciling discrepancies among earlier observational and simulation studies, and provide new insight into how compression and velocity shear modulate solar wind turbulence near CIRs.

\end{abstract}



\section{Introduction} \label{sec:intro}

Solar wind turbulence is a fundamental phenomenon in space, offering valuable insights into energy transfer, particle acceleration, and plasma heating processes across the heliosphere. Moreover, as a magnetized, collisionless plasma, the solar wind serves as an ideal laboratory for investigating the nature of plasma turbulence. Therefore, understanding turbulence in the solar wind is essential for advancing plasma physics, heliospheric science, and astrophysics.

Fully developed turbulence is often described as ``unpredictable'', reflecting its chaotic behavior.
This behavior is evident across a broad range of timescales, from seconds to months, exhibiting self-similarity \citep{bruno_solar_2013}. Some solar winds are characterized by the high Alfv\'enicity, marked by a strong correlation between the velocity and magnetic field fluctuations, implying Alfv\'en waves propagating away from the sun \citep{belcher_large-amplitude_1971,velli_waves_1991}. 
The properties of solar wind turbulence differ notably between slow and fast wind streams. 
Fast solar wind, typically originating from large coronal holes at high solar latitudes \citep{phillips_ulysses_1995, mccomas_ulysses_1998}, is relatively uniform with higher Alfv\'enicity, indicating its magnetic field and velocity fluctuations are closely aligned. 
In contrast, slow solar wind, which generally emerges from near the solar equator and active regions \citep{phillips_ulysses_1995, mccomas_ulysses_1998}, exhibits greater variability in density, composition, and velocity, and often displays relatively lower Alfv\'enicity \citep{tu_dependence_1990}. 
Compared to fast wind, slow wind is typically slower, denser and has a lower proton temperature on average \citep{mccomas_solar_2000,neugebauer_role_1976, wilson_iii_statistical_2018}.

\begin{figure*}[ht!]
\includegraphics[scale=1]{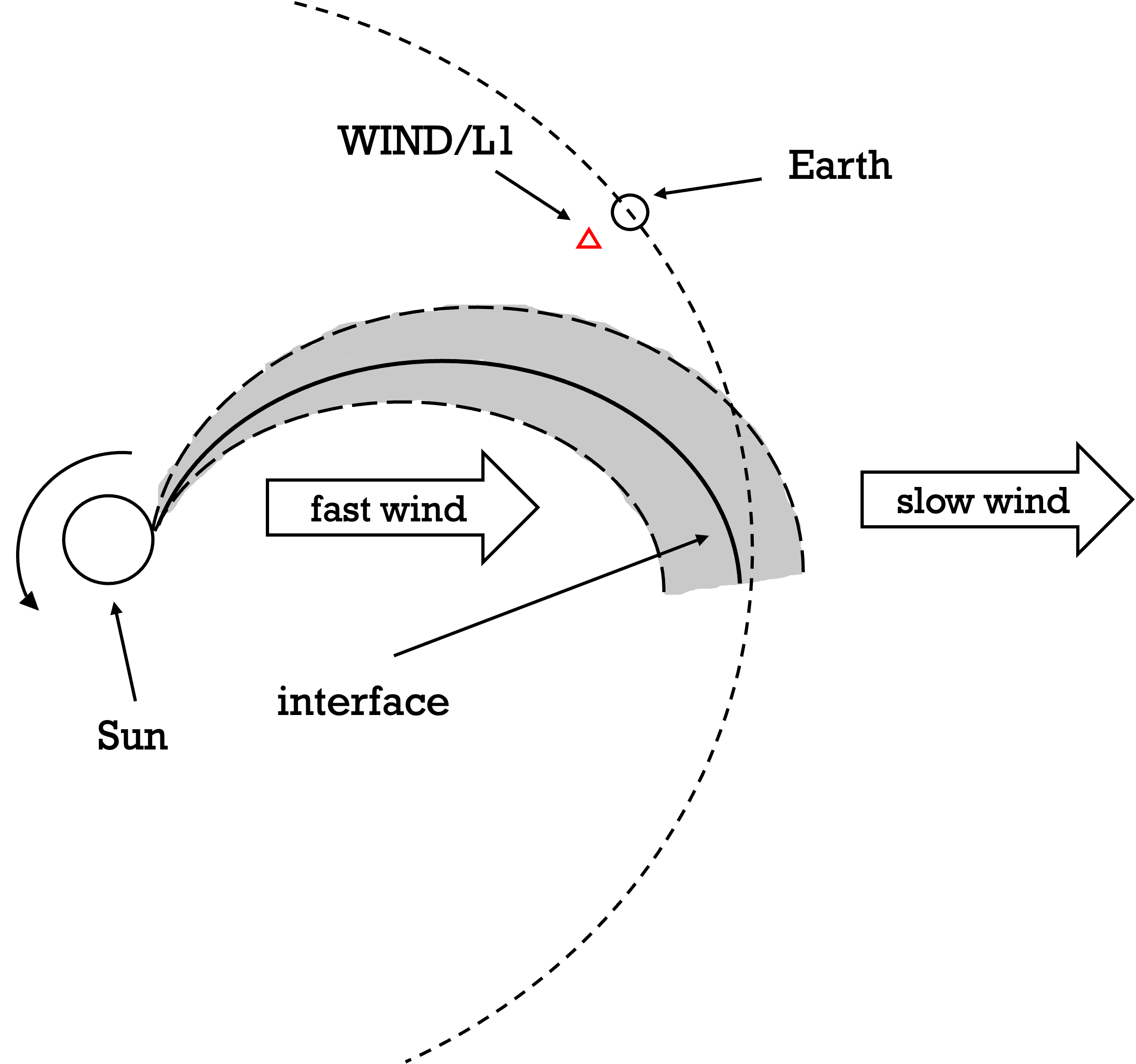}
\caption{A general 2D structure of the CIR(grey region). The position of WIND spacecraft is plotted with a red triangle. This figure is inspired by \cite{borovsky_solar_2010}
\label{fig:CIR_struc}}
\end{figure*}

These differences become particularly pronounced within Corotating Interaction Regions (CIRs), also known as Stream Interaction Regions (SIRs), which form as fast solar wind streams collide with slower-moving solar wind ahead of them \citep{gosling_compressions_1972, gosling_solar_1978, gosling_formation_1999} (see Fig. \ref{fig:CIR_struc}). This collision compresses plasma, enhances magnetic fields, and generates turbulent structures within CIRs, creating a unique environment where fast and slow wind properties converge. 
At the center of the CIR lies the stream interface, which serves as the boundary between the slow and fast solar wind \citep{gosling_solar_1978}. Observations of stream interfaces were first reported in 1971 using Mariner 5 data, and it was noticed that the compression region near the interface may contain inwardly propagating or non-Alfv\'enic waves \citep{belcher_large-amplitude_1971}. 
Later in 1978, \citet{gosling_solar_1978} examined the properties of interaction regions in detail, performing Superposed Epoch Analysis for 23 discontinuous stream interfaces. They noted both the electron and proton temperature increased at the interface. \citet{jian_properties_2006} carried out a comprehensive survey of CIRs/SIRs at 1 Astronomical Unit (AU) based on in situ observations by WIND and ACE. By statistically analyzing the shock association, they revealed SIRs with forward shocks are in majority compared to those with reverse shocks, and the shock association rate is larger than the rate at 0.72 AU \citep{lindsay_sources_1994}. 
Later in 2019, they conducted another statistical analysis on SIRs and their relationship with Heliospheric Current Sheets (HCS), revealing that the SIRs with current sheets generally have slightly lower minimum speed and maximum speed, but higher peak in proton density, magnetic field, total pressure and dynamic pressure.\citep{jian_solar_2019}. 

More recently, the Parker Solar Probe (PSP) has provided an unprecedented opportunity to study the evolution and characteristics of CIRs in regions much closer to the Sun. Several CIRs were observed and analyzed during PSP’s early orbits, with a focus on their relationship to energetic particles \citep{mccomas_probing_2019, cohen_energetic_2020, desai_properties_2020, Joyce_energetic_2020}. 
To support future investigations of CIR evolution, \citet{allen_living_2021} created a catalog of CIRs and SIRs observed by both PSP and other heliophysics missions at 1 AU. 
Later, \citet{allen_radial_2021} reported a radial-aligned CIR event observed by PSP at 0.5 AU and the Solar Terrestrial Relations Observatory-Ahead(STEREO-A) at 1 AU. 
They found that while PSP recorded lower suprathermal ion intensities, it also observed enhancements persisting over a longer duration and wider longitudinal extent compared to STEREO-A. 
This was attributed to the magnetic field line geometry.

Previous study have shown that comparing the two streams near the CIRs, the fast stream is highly Alfv\'enic while in the slow stream the domination of outward mode is weaker \citep{tu_dependence_1990}. However, the characteristics of turbulence in the transition region from slow stream to fast stream have not yet reached a widely-accepted consensus. \citet{borovsky_solar_2010} performed a superposed epoch analysis on 27 CIRs using ACE and WIND data. They observed a dip of Alfv\'en ratio at the interface, indicating a decreased ratio of kinetic fluctuation energy to magnetic fluctuation energy. They also showed a spike of ratio of Elsasser amplitudes at the interface, indicating an increased ratio of outward Alfv\'en wave energy to inward Alfv\'en wave energy. In contrast, observations from Helios 2 at approximately 0.3 AU revealed a drop in normalized cross helicity ($\sigma_c$)-which quantifies the ratio of outward to inward Alfv\'en wave energy-within a velocity shear zone \citep{roberts_velocity_1992}. 
Similarly, \citet{shi_alfvenic_2021}, reported a decrease in $\sigma_c$ and a peak in normalized residual energy ($\sigma_r$) near a velocity shear region at 20 solar radii using PSP data. Their simulation of fast–slow stream interactions using an expanding-box model showed results consistent with these observations \citep{shi_propagation_2020,shi_influence_2022}. Obviously, a further investigation is necessary to better understand the properties of solar wind turbulence within CIRs.

In this paper, we analyze the turbulence properties within CIRs using WIND data and Superposed Epoch Analysis(SEA). We employ Minimum Variance Analysis(MVA)\citep{sonnerup_magnetopause_1967, sonnerup_minimum_1998} to determine the stream interface of each CIR and apply wavelet transform to evaluate the fluctuation energies.
Our findings reveal that all forms of fluctuation energy are affected to varying degrees by the compression between the fast and slow solar wind. A drop in $\sigma_c$ near the stream interface is observed in different frequency ranges, which is a symbol of increased inward Alfv\'en wave energy caused by the compression. In low frequency part, a strong peak of $\sigma_r$ is observed near the interface, which is influenced by bringing the large-scale velocity shear into the velocity fluctuation energy. In mid and high frequency ranges, as frequency decreases, $\sigma_c$ shows an increase in fast stream while $\sigma_r$ shows a decrease over the whole data range. In the discussion section, we compare our findings with previous studies, offering possible explanations for the inconsistent results in these research about $\sigma_c$ and $\sigma_r$. 

\section{Data Selection and Processing} \label{sec:style}
\subsection{Data Selection} \label{subsec:select}
The dataset utilized in this study comprises merged magnetic field and plasma measurements from the WIND spacecraft, which operates at the L1 Lagrangian point. The magnetic field data is from the Wind Magnetic Field Investigation (MFI) \citep{lepping_wind_1995}, level \textbf{h0}, with a cadence of 3 s. The proton density, solar wind velocity and temperature measurements are obtained from WIND 3D Plasma Analyzer (3DP) electrostatic analyzers \citep{lin_three-dimensional_1995} 
with a time interval of 3 s, while the electron data are from the Solar Wind Experiment (SWE) electron instruments \citep{ogilvie_swe_1995}, with a
cadence of 6–12 s. The measurements are resampled to a cadence of 12 s, which is sufficient for the analysis of MHD turbulence at 1 AU. Data gaps are filled using linear interpolation, and events with a data gap ratio exceeding 20$\%$ are excluded from the analysis.

For the SEA of solar wind turbulence at 1 AU, we collected the CIRs occurring from 2003 to 2021. 
The criteria for the identification of CIR include: (1)They have a clear shear zone where velocity increases sharply within 24 hours, (2)for each event, the difference between the maximum and minimum velocity must exceed $100km/s$, (3)the wind speed in the fast wind side is higher than the slow wind side for at least 24 hours. Interplanetary Coronal Mass Ejections (ICMEs) are removed based on the presence of a shock at the interface between fast and slow streams and extremely high solar wind velocity ($\geq800km/s$) in the fast wind side. In addition, an existing ICME catalog is consulted to ensure that no events were mistakenly included or excluded.\citep{richardson_near-earth_2024} After careful selection, we obtained a list of 237 events (examples available in Table \ref{tab:list}).
We noticed that a large number of these events are accompanied with HCS, so we conducted a statistical analysis to check the distribution of CIRs with and without HCS, which will be discussed in Section \ref{sec:discussion}.

\begin{table}[h!]
\centering
\caption{The first seven L1 CIR events in our list.}
\label{tab:list}
\begin{tabular}{ccc}
\hline
Index & Time:stream interface & Time: HCS to SI (day) \\
\hline
1 & 2003-01-19 07:46:49.404 & -1.771530 \\
2 & 2003-02-04 05:36:24.294 &  \\
3 & 2003-05-28 01:45:25.959 &  \\
4 & 2003-06-26 15:34:41.932 & -0.085812 \\
5 & 2003-07-03 21:37:52.134 &  \\
6 & 2003-07-11 21:37:02.394 &  \\
7 & 2003-07-26 21:37:03.541 & -0.394596 \\
\hline
\end{tabular}
\tablecomments{This is an example of CIRs in our list. The blank time indicates no HCS association. The whole list can be found in the HTML version full article.}
\end{table}

MVA is used to identify the stream interface between the fast and slow wind—referred to as the zero epoch in SEA—based on the assumption that the velocity changes most rapidly at this boundary. Originally developed to estimate the normal direction to transition layers in space plasma from spacecraft data \citep{sonnerup_magnetopause_1967}, MVA determines the directions along which a vector field exhibits maximum and minimum variance. In this study, it allows us to estimate a unit vector $\boldsymbol{\hat{n}}$ along which the velocity variance is largest. We then define the data point with the maximum velocity gradient along this direction as the zero epoch. To be specific, starting from a velocity field $\{\boldsymbol{v}^{(m)}\}(m=1,2,3...,M) $ and after the determination of direction $\boldsymbol{\hat{n}}$ corresponding to the MVA, we get a time series of velocities projected on this direction $\{\boldsymbol{v}^{(m)}\cdot\boldsymbol{\hat{n}}\}(m=1,2,3...,M) $. To reduce the influence of fluctuations, these data are then averaged using a two-hour running window. The zero epoch is finally identified as the point where the averaged velocity exhibits the largest gradient between two consecutive data points.

\begin{figure*}[ht!]
\plotone{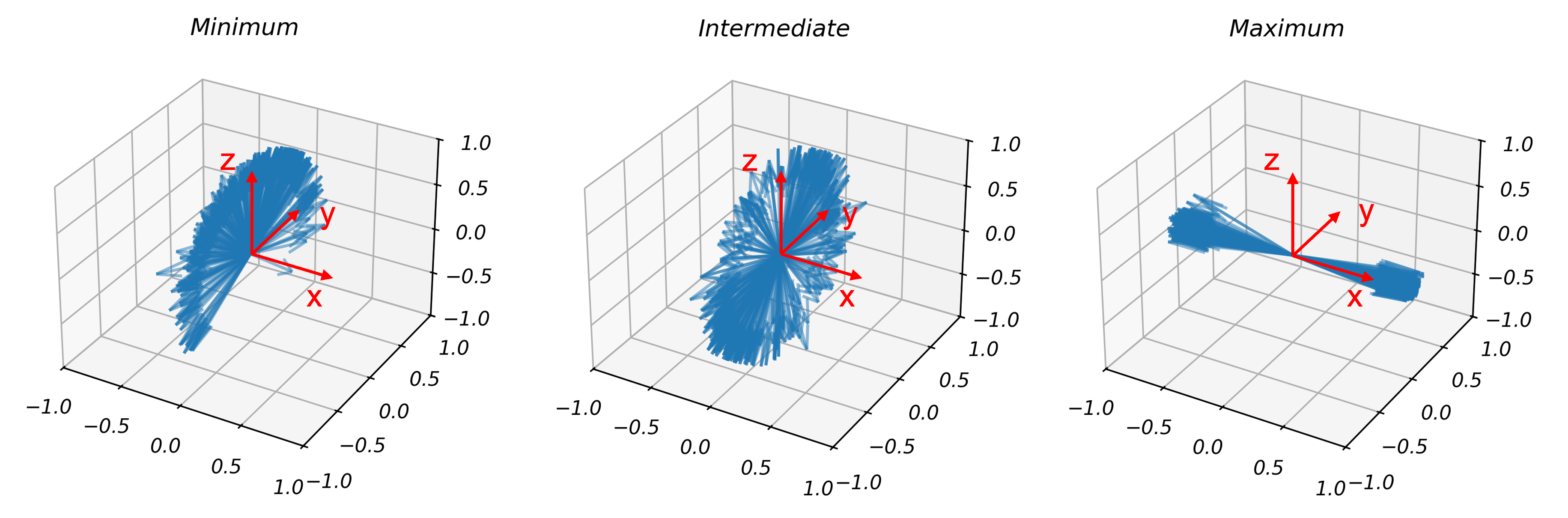}
\caption{The distributions of the unit vectors determined by MVA in 3D GSE coordinate system, with x pointing towards the Sun from the Earth. From left to right panels, the vectors are with minimum, intermediate and maximum variance, respectively.
\label{fig:angle_distri}}
\end{figure*}

By applying MVA to the velocity field we obtain the direction of compression. However, this direction does not correctly represent the normal vector of the CIR transition layer. In the compressed solar wind, the condition $\nabla \cdot \boldsymbol{v} = 0$ no longer strictly holds. Nonetheless, under the assumption that the CIR is a relatively stationary structure, the continuity equation shows  $\nabla \cdot (\rho \boldsymbol{v}) = 0$ is still valid. The jump condition at the CIR transition layer implies that $\rho \boldsymbol{v} \cdot \boldsymbol{n}$ remains approximately constant across the layer, where $\boldsymbol{n}$ denotes the normal vector. Therefore, the direction of minimum variance of $\rho \boldsymbol{v}$ can be regarded as the direction normal to the CIR interface. 
Through the application of MVA to the mass flux $\rho \boldsymbol{v}$, we obtain three orthogonal vectors for each event, corresponding to the directions of minimum, intermediate, and maximum variance. Fig. \ref{fig:angle_distri} shows the distribution of these vectors in a 3D Geocentric Solar Ecliptic (GSE) coordinate system. The minimum variance vectors-normal vectors-lie predominantly within the y–z plane, while the maximum variance vectors-the parallel ones-are closely aligned with the x-direction. These indicate the angle between the interface and the x-direction is small, rather than the Parker spiral angle of approximately $45^{\circ}$ mentioned in some previous studies \citep{richardson_major_2006, borovsky_solar_2010}.
However, there exists a high possibility that our results are affected by systematic errors. Considering the solar wind velocity and mass flux are much smaller in the y- and z-directions than in the x-direction, the maximum variance vector naturally tends to align with the x-axis, while the minimum variance direction typically lies within the y–z plane. 
Quantifying the impact of these systematic effects is difficult; therefore, we present these results here as a reference rather than a definitive conclusion.

\subsection{Preliminary Analysis} \label{subsec:pre}
\begin{figure*}[ht!]
\plotone{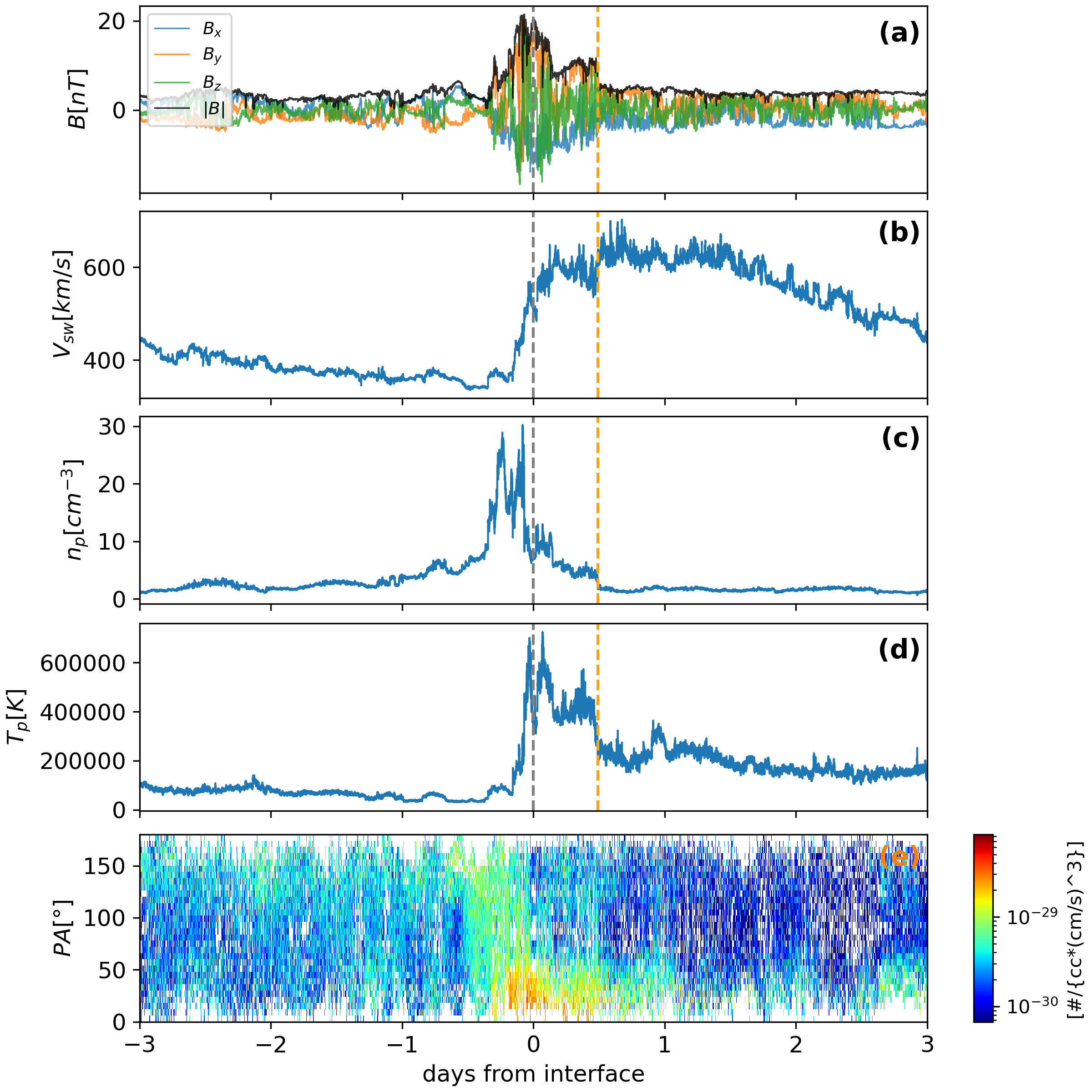}
\caption{A typical CIR on August 7, 2006, from WIND data. The zero epoch time is the stream interface(vertical dashed line). \textbf{(a)} The three magnetic field components (blue, orange and green) in the GSE coordinate system, and the black curve is the magnitude. \textbf{(b)} The solar wind speed. \textbf{(c)} The proton density. \textbf{(d)} The proton temperature. \textbf{(e)} The electron pitch angle distribution (PAD) from channel ``f\_pitch\_E07'' (290.1 eV) of SWE, revealing a distinct current sheet near the zero epoch. The cadence of all panels is 12 seconds. The orange dashed line aligns with a reverse shock at 0.5 day from interface.
\label{fig:profile}}
\end{figure*}

A representative example of a CIR  from our list is shown in Fig. \ref{fig:profile}. 
The data extends from 3 days prior to the passage of the stream interface to 3 days afterward. The gray dashed line, marking the zero epoch, aligns precisely with the interface between the slow and fast stream, confirming the validity of using MVA to determine the zero epoch. 
Panels (a) and (b) clearly show that the magnetic field and solar wind velocity have more intense fluctuations in the fast wind side. 
In panel (c), the proton density peak appears on the slow wind side, just before the interface, a feature that is also reflected in the subsequent superposed analysis. 
Approximately 12 hours after the interface-marked by the orange dashed line-a sudden increase in $V_{sw}$ is observed, accompanied by abrupt decreases in $N_p$ and $T_p$, indicating the presence of a reverse shock.
This CIR is accompanied by a HCS at the zero epoch, as clearly illustrated in the fifth panel, where the peak of electron pitch angle distribution (PAD) shifts from around 180$^{\circ}$ to around 0$^{\circ}$ near the interface. 
In our research, we categorized the CIRs based on the presence or absence of HCS, and a more detailed analysis will be discussed in Section \ref{sec:discussion}. 

\begin{figure*}[ht!]
\plotone{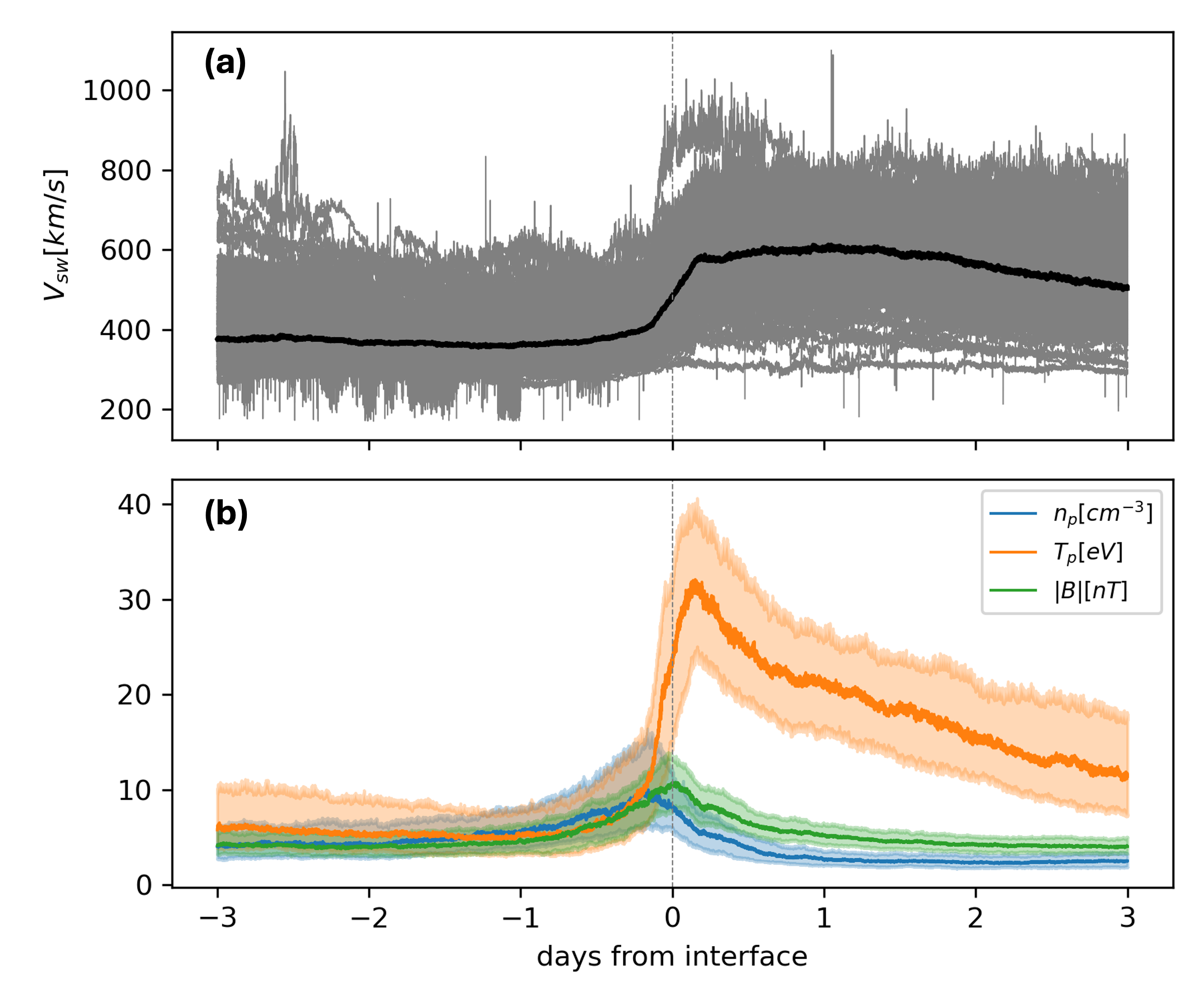}
\caption{\textbf{(a)} The superposed median of solar wind velocity. The gray lines are of the individual events. The black solid line is the median velocity of all events. \textbf{(b)} The superposed median of proton density(blue), proton temperature(orange), and magnetic field amplitude(green). The light-colored areas show the data between 25\% and 75\% of all events for reference.
\label{fig:vsw_den_temp}}
\end{figure*}

To acquire a comprehensive understanding of the common characteristics of CIRs, we calculated the superposed median of flow velocity and several plasma parameters of all events in our list, as shown in Fig. \ref{fig:vsw_den_temp}. The top panel includes the solar wind velocity of individual events (gray) and the median flow velocity (black). The rise of flow velocity occurs with the passage of stream interface. Panel (b) shows the SEA results for proton density, temperature, and magnetic field strength, where a distinct peak in proton density is clearly visible on the slow wind side, just before the stream interface \citep{gosling_solar_1978,richter_superposed_1986,borovsky_solar_2010}. We note that there is a bump of the magnetic field amplitude at the zero-epoch with elevated field strength on both sides near the interface \citep{gosling_solar_1978, richter_superposed_1986,borovsky_solar_2010}. Also note the proton temperature $T_p$ is low in the slow stream and sharply increases near the interface, peaking in the fast wind side and slowly falling down \citep{gosling_solar_1978,borovsky_solar_2010}. These different peak positions of $|B|$, $n_p$ and $T_p$ are also showed in simulation by \cite{shi_influence_2022}. Given the higher density and lower temperature of the slow wind, it is expected that compression causes the density peak to lie closer to the denser slow wind, while the temperature peak appears nearer to the hotter fast wind.

\section{Turbulence Analysis} \label{sec:floats}
In our study of MHD turbulence, we focus on two critical quantities, the normalized cross helicity
\begin{equation}
\sigma_c=\frac{E_o-E_i}{E_o+E_i}
\label{eq:sigmac}
\end{equation}
which measures the ratio of outward Alfv\'en wave energy to inward Alfv\'en wave energy, and the normalized residual energy
\begin{equation}
\sigma_r=\frac{E_v-E_b}{E_v+E_b}
\label{eq:sigmar}
\end{equation}
which measures the ratio of kinetic fluctuation energy to magnetic fluctuation energy. For Alfv\'en wave energy, we use two Els\"asser variables
\begin{equation}
\mathbf{z}_{o,i}=\mathbf{v}\mp sign(B_{r0})\frac{\mathbf{B}}{\sqrt{\mu_0\rho}}
\end{equation}
which correspond to the outward and inward Alfv\'en waves. The subscripts ``o'' and ``i'' represent ``outward'' and ``inward'', respectively. As for the kinetic and magnetic energy, we use 
\begin{equation}
\mathbf{v}_a=\frac{\mathbf{B}}{\sqrt{\mu_0\rho}}
\end{equation}
for the estimation of magnetic fluctuation energy, and original $\mathbf{v}$ for the kinetic fluctuation. 
In Eq. (4), the density $\rho$ is averaged in 7-minute running window. Now the expression of energies can be written as $E_o=|\mathbf{z}_o|^2$, $E_i=|\mathbf{z}_i|^2$, $E_k=|\mathbf{v}|^2$, $E_b=|\mathbf{v}_a|^2$, where the $\mathbf{z}_o$, $\mathbf{z}_i$, $\mathbf{v}_a$ and $\mathbf{v}$ stand for the fluctuations. Common methods for estimating the fluctuation include using running window average to subtract the background field \citep{borovsky_solar_2010}, and applying Fourier transform or wavelet transform as a bandwidth filter to select the data within a specific frequency range \citep{grappin_alfvenic_1991,  shi_alfvenic_2021, sioulas_magnetic_2023, huang_new_2023}. 
In this study, we employ wavelet transform to obtain the energy in turbulence, which offers the advantage of providing 2-D power spectrogram with information across both time and frequency domains, eliminating the need to define a specific time window size. Additionally, wavelet transform effectively mitigates the challenge of analyzing fluctuations within an inhomogeneous background field, reducing the influence of the rapid transition from slow wind to fast wind. However, this influence cannot be completely eliminated and becomes significant at very low frequencies, as will be discussed later in this section.

\begin{figure*}[ht!]
\plotone{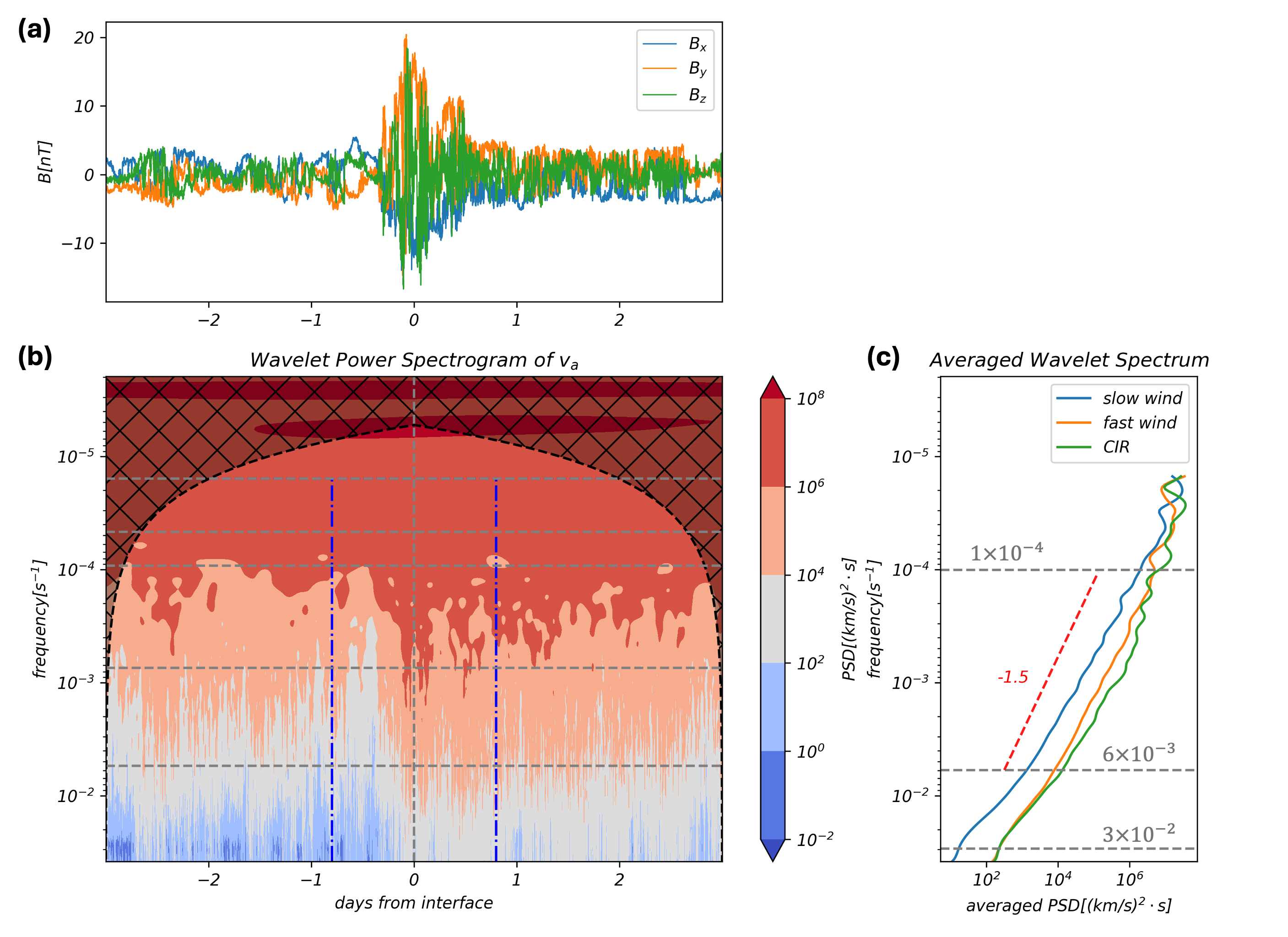}
\caption{Using wavelet transform to get the magnetic power spectrum of one CIR on August 7, 2006. Panel (a) and (b) share the x axis, while panel (b) and (c) share the y axis. \textbf{(a)} The magnetic field components in GSE coordinate with a cadence of 12 seconds. \textbf{(b)} The power spectrogram of magnetic fluctuation, which is the quadratic sum of the power spectral density in three components. The shaded area is where the results from the wavelet transform become unreliable, as it lies within the Cone of Influence (CoI). Horizontal dashed lines divide the frequencies into five regions for calculating $\sigma_c$ and $\sigma_r$ across different frequency ranges. The vertical gray dashed line marks the zero epoch, while the vertical blue dash-dotted lines at -0.8 and 0.8 indicate the approximate boundaries between the slow wind, CIR region, and fast wind. \textbf{(c)} Averaged wavelet spectra in three regions separated by the blue dash-dotted lines in panel (b). The gray dashed lines are the boundaries of the inertial range (from $1\times10^{-4}~s^{-1}$ to $6\times10^{-3}~s^{-1}$) and the dissipation range (from $6\times10^{-3}~s^{-1}$ to $3\times10^{-2}~s^{-1}$).
\label{fig:wavelet}}
\end{figure*}

The method we used is continuous wavelet transforms (CWT) implemented in the Python package PyWavelet \citep{lee_pywavelets_2019}. 
The mother function is Morlet wavelet, and the cadence is 12 seconds. For each variable ($\mathbf{z}_o$, $\mathbf{z}_i$, $\mathbf{v}_a$ and $\mathbf{v}$) in one case, we first apply CWT to all three components in the whole data range, from 3 days before the passage of interface to 3 days after that.
Then we calculate the quadratic sum of three components, which is the trace spectrogram of this variable. One example of the result is shown in Fig. \ref{fig:wavelet}, the spectrogram of $\mathbf{v}_a$ in a CIR on August 7, 2006. Panel (a) shows the resampled magnetic field data, which is the same as that presented in Fig. \ref{fig:profile}. These three components are normalized to $v_{a,x}$, $v_{a,y}$, and $v_{a,z}$ using a 7-minute averaged density. After applying the wavelet transform, the resulting power spectrogram is shown in panel (b). Different colors mark the power spectrum density (PSD). The slow wind, CIR region, and fast wind are separated by the blue dash-dotted lines, which mark approximate boundaries at -0.8 and 0.8 days from the interface. Notably, the power spectrogram shows higher values on the fast wind side compared to the slow wind, with the highest values occurring in the CIR region near the interface. 
This becomes more evident in the time-averaged spectra of the three regions shown in panel (c): the fast wind curve is clearly higher than that of the slow wind, while the green curve representing the CIR region is slightly higher than the fast wind.
This difference aligns with expectations, as the fast stream carries more fluctuation energy than the slow stream, and the CIR region contains even greater energy due to the compression.
To conduct a further analysis of the power spectra, the spectral index is calculated by linear fitting in the inertial range and dissipation range. The two ranges are determined by the gray dashed lines in panel (c), with inertial range from $1\times10^{-4}~s^{-1}$ to $6\times10^{-3}~s^{-1}$ and dissipation range from $6\times10^{-3}~s^{-1}$ to $3\times10^{-2}~s^{-1}$. The inertial range indices for the slow wind, fast wind, and CIR region are -1.77, -1.60, and -1.48, respectively, while the corresponding dissipation range indices are -2.93, -2.36, and -2.68. The red dashed line with a slope of -1.5 is plotted in the inertial range for reference.

\begin{figure*}[ht!]
\plotone{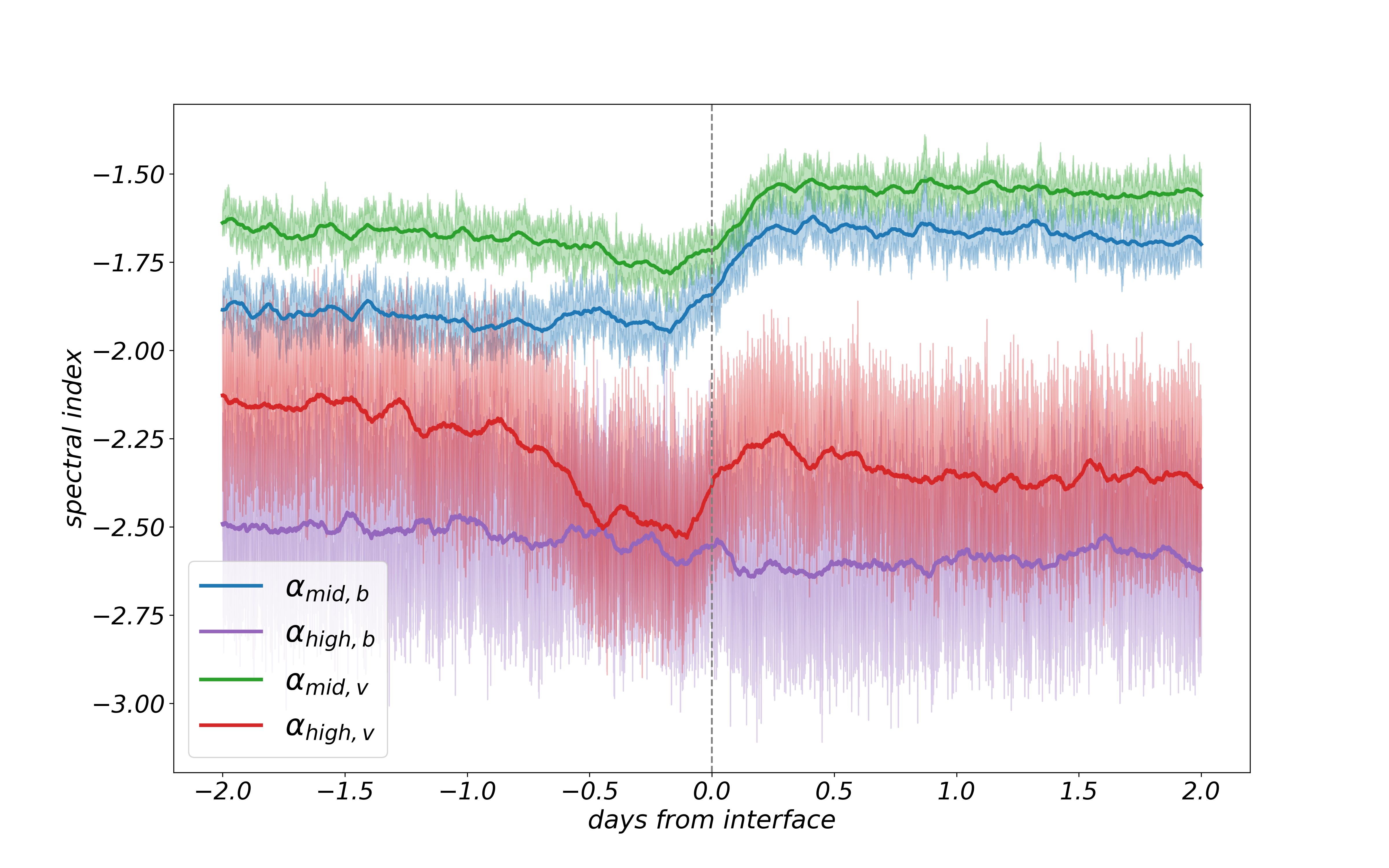}
\caption{The superposed median of the spectral indices of magnetic field and velocity in both the inertial and dissipation ranges. The light-colored areas indicate the data between the 45\% and 55\% of all events, provided for reference. The subscript ‘mid’ refers to the mid-frequency (inertial) range, ‘high’ to the high-frequency (dissipation) range, ‘b’ to the normalized magnetic field $v_a$, and ‘v’ to the velocity.
\label{fig:index}}
\end{figure*}

We further calculated the spectral indices of magnetic field and velocity fluctuations at various time points from -2 to +2 days relative to the interface. The superposed medians of all events are shown in Fig. \ref{fig:index}, with light-colored areas indicating the range between 45th and 55th percentiles for errors. Both in the inertial and dissipation ranges, the index of magnetic field-blue and purple curves-tends to be steeper than the velocity spectral index-green and red curves. In inertial range, magnetic and velocity spectral indices show similar trends, smoothly shifting from steeper slow wind side to shallower fast wind side. The blue curve increases from around -1.85 to around -1.6, which is consistent with some previous works \citep{borovsky_solar_2010, alonso_guzman_superposed_2025, ghanbari_galactic_2019, chen_residual_2013}. In the dissipation range, the two curves exhibit similar behavior, except for a dip in the velocity spectral index just before the stream interface. This could be due to enhanced dissipation of velocity fluctuation energy caused by compression, or it could result from the influence of the strong velocity shear. However, it is important to note that the error bands in the dissipation range are significantly larger than those in the inertial range, indicating that the dissipation spectral indices are less reliable.

\begin{figure*}[ht!]
\plotone{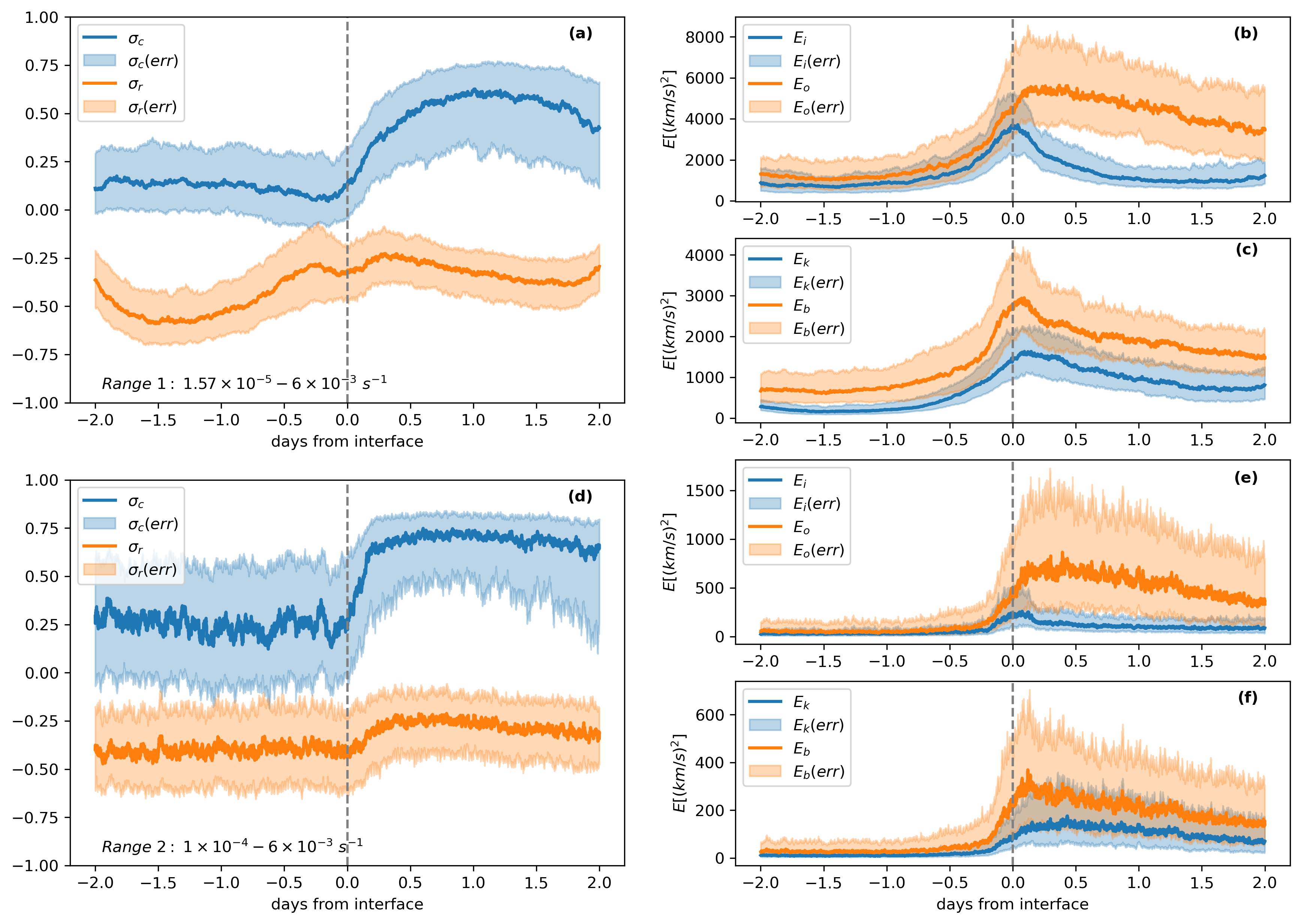}
\caption{\textbf{(a)} Superposed median of $\sigma_c$ and $\sigma_r$ within Range 1. The time range here is $\pm$2 days from the passage of stream interface. \textbf{(b)} Superposed median of inward Alfv\'en energy $E_i$(blue) and outward Alfv\'en energy $E_o$(orange) within Range 1. \textbf{(c)} Superposed median of kinetic fluctuation energy $E_k$(blue) and magnetic fluctuation energy $E_b$(orange) within Range 1. \textbf{(d)}, \textbf{(e)}, and \textbf{(f)} are analogous to \textbf{(a)}, \textbf{(b)}, and \textbf{(c)} but use data from Range 2 instead of Range 1. All the light-colored areas show the data between 25\% and 75\% of all data at each time point.}
\label{fig:final}
\end{figure*}

Turbulence energies $E_k$, $E_b$, $E_i$ and $E_o$ are estimated by integrating the corresponding power spectrogram along the frequency axis at each time moment. The quantities $\sigma_c$ and $\sigma_r$ are then calculated using equations \ref{eq:sigmac} and \ref{eq:sigmar}. By selecting different frequency ranges for the integration, we can obtain the fluctuation energies and turbulence properties across various time scales. 
We first choose two ranges: Range 1 is of almost all frequencies, from $1.57\times10^{-5}~s^{-1}$ to $6\times10^{-3}~s^{-1}$, while Range 2 is without the low frequency part, from $1\times10^{-4}~s^{-1}$ to $6\times10^{-3}~s^{-1}$. 
The SEA results for these two ranges are shown in Fig. \ref{fig:final}, with panels (a), (b) and (c) corresponding to range 1 and panels (d), (e) and (f) corresponding to range 2. 
To eliminate the boundary effects of CWT, the time range here is $\pm2$ days from the interface instead of 3 days. In panel (a), $\sigma_c$ increases from approximately 0.1 to 0.6 from slow wind to fast wind, and $\sigma_r$ increases from around -0.6 to -0.55, suggesting enhanced outward dominance and weakened magnetic dominance in the fast solar wind compared with the slow stream. 
To better explain the changes of $\sigma_c$ and $\sigma_r$, superposed medians of fluctuation energy are plotted in panel (b) and (c). 
Notably, $E_o$ is significantly higher in the fast wind side compared with the slow wind side, whereas the change of $E_i$ from slow wind to fast wind is not very obvious.
This suggests that the increase in $\sigma_c$ is primarily driven by the higher outward Alfv\'en wave energy.
However, a distinct peak of $E_i$ occurs at the stream interface, indicating the enhanced inward Alfv\'en wave energy. 
This enhancement leads to a drop of $\sigma_c$ near the interface, which is consistent with our previous observation \citep{shi_alfvenic_2021} and simulation \citep{shi_influence_2022}. 

\begin{figure*}[ht!]
\plotone{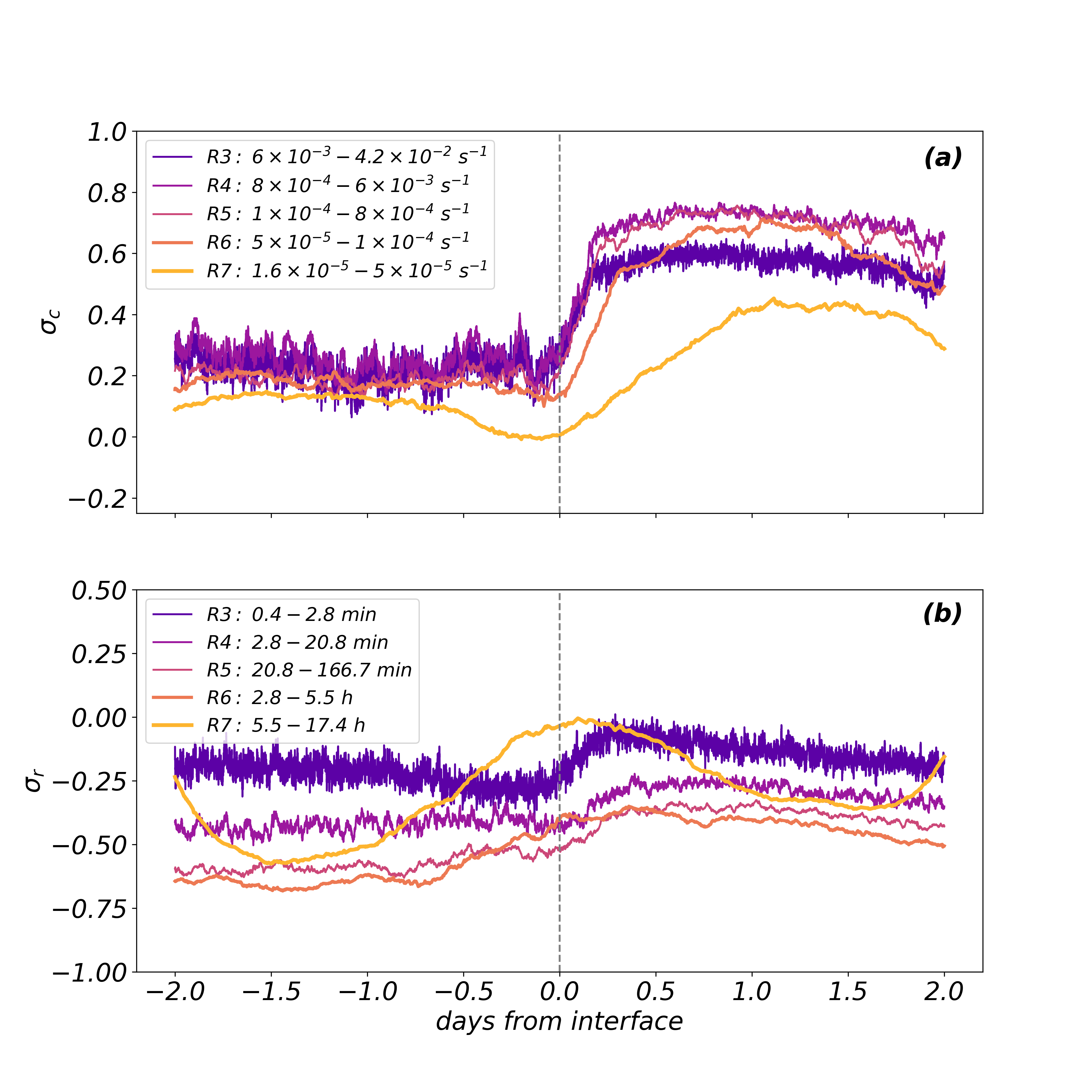}
\caption{Superposed median of $\sigma_c$ and $\sigma_r$ in different frequency ranges. ``R'' in the legend stands for ``Range''. The time cadence is 12s. Legends indicate frequency ranges in panel \textbf{(a)}, and the corresponding time scales in panel \textbf{(b)}.
\label{fig:sigmas}}
\end{figure*}

The behavior of $\sigma_r$ near the zero epoch is more complex. An overall increase is observed near the interface, with a dip exactly at the zero epoch and two small peaks on either side, forming a ``peak-valley-peak'' structure. 
However, these features disappear in panel (d) when the low-frequency range is excluded. 
In this case, the drop in $\sigma_c$ due to enhanced $E_o$ remains visible, while $\sigma_r$ exhibits a smoother transition from slow to fast wind. Additionally, in panel (f), the magnetic and kinetic fluctuation energies no longer display the pronounced near-interface peaks seen in panel (c). 
One possible explanation is that very low-frequency structures—such as the velocity shear between slow and fast wind—have a strong influence on the estimated fluctuation energy, although they are not turbulence or fluctuations. To test this assumption, we selected five narrower frequency ranges to calculate $\sigma_c$ and $\sigma_r$, as indicated by the gray dashed lines in panel (b) of Fig. \ref{fig:wavelet}. The boundaries are as follows: Range 3 is from $6\times10^{-3}~s^{-1}$ to $4.2\times10^{-2}~s^{-1}$, Range 4 is from $8\times10^{-4}~s^{-1}$ to $6\times10^{-3}~s^{-1}$, Range 5 is from $1\times10^{-4}~s^{-1}$ to $8\times10^{-4}~s^{-1}$, Range 6 is from $5\times10^{-5}~s^{-1}$ to $1\times10^{-4}~s^{-1}$, Range 7 is from $1.6\times10^{-5}~s^{-1}$ to $5\times10^{-5}~s^{-1}$. The SEA results are shown in Fig. \ref{fig:sigmas} with gradient colors. 
From Range 3 to Range 6, the $\sigma_c$ curves are quite similar on the slow wind side, whereas the curve of Range 7 is relatively lower. On the fast wind side, however, the $\sigma_c$ value shows an increase-decrease trend from Range 3 to Range 7 as the frequency decreases, suggesting that the dominance of outward Alfv\'en wave energy is strongest around Ranges 4 and 5. At the high-frequency end (Range 3), the scales approach the kinetic regime, while at the low-frequency end (Range 7), the analysis may encompass fluctuations across different wind streams. Both factors could contribute to the lower $\sigma_c$ observed on the fast wind side.
In addition, note that the $\sigma_c$ shows a stronger dip within CIR region in range 6 and 7, while the overall trend remains the same. For $\sigma_r$, it is evident that as the frequency decreases from Range 3 to Range 5, $\sigma_r$ shows an overall decline, indicating a stronger dominance of magnetic fluctuations in the lower-frequency range. However, at very large time scales—corresponding to Ranges 6 and 7—a significant rise in $\sigma_r$ emerges near the stream interface, with the increase being even more pronounced in Range 7 than in Range 6. We note that the smallest time scale in Range 6 is approximately 2.8 hours, which is very close to the scale of the velocity shear region. It is possible that in a frequency range encompassing these large time scales, the estimated velocity fluctuation energy is heavily influenced by the velocity shear, leading to the observed increase of $\sigma_r$ within the shear region. We believe this can explain the overall increase and the peaks of $\sigma_r$ in the panel (a) of Fig. \ref{fig:final}. The ``valley'' in the ``peak-valley-peak'' structure may result from a mixing effect when calculating fluctuation energy over a wide frequency range, as this feature is not obvious in the narrow frequency bands of Ranges 3 to 7. 

\section{Discussion} \label{sec:discussion}
Our analysis highlights the impact of compression  on the turbulence properties within CIRs, and indicates the overestimation of velocity turbulence caused by the velocity shear. The influence of compression is evident across various frequency ranges, characterized by enhanced fluctuation energy—particularly the increase in inward Alfv\'en wave energy.
This increase in $E_i$ leads to a dip of $\sigma_c$ at the stream interface.
The velocity shear plays an important role in the low frequency ranges, where the enhancement of $E_k$ results in a pronounced peak of $\sigma_r$ near the interface.
Additionally, the superposed data on either side of the interface offers a valuable comparison between the fast and slow wind at 1 AU. One notable result is that the kinetic fluctuation energy, magnetic fluctuation energy and outward Alfv\'en wave energy are all substantially higher in the fast wind compared to the slow wind, while the inward Alfv\'en wave energy remains relatively constant in both regimes.

Our figures for $\sigma_c$ and $\sigma_r$ are not fully consistent with the results reported in \cite{borovsky_solar_2010}. 
(1)In their study, the ratio of Elsasser amplitudes is used to quantify the relative amplitude of outward to inward Alfv\'en waves, which is analogous to $\sigma_c$. However, they did not observe a drop at the interface as shown in Fig. \ref{fig:final}. Additionally, in our results, the inward Alfv\'en wave energy shows an apparent rise at the interface, which is not observed in their figure. 
(2)As the equivalent of $\sigma_r$, they used Alfv\'en Ratio $r_A=\frac{(\delta v/v_A)^2}{(\delta B/B)^2}$ to measure the relative amplitude of velocity fluctuation energy to magnetic fluctuation energy. In their results, $r_A$ exhibits a distinct drop at the stream interface, rather than the rise for low frequencies or the smooth transition for mid- and high frequencies observed in our study (see $\sigma_r$ in Fig. \ref{fig:sigmas}). In addition, the $r_A$ in their figure 11 is approximately 0.8, which corresponds to around -0.1 when converted to $\sigma_r$. 
Our superposed median of $\sigma_r$ is between -0.6 and -0.2 in slow wind and within -0.5 and -0.1 in fast wind, and this value is supported by several previous statistical analysis \citep{damicis_origin_2015,sioulas_magnetic_2023, chen_residual_2013}. 
We believe the inconsistency primarily arises from differences in the methods used to calculate the energy in fluctuations. 
In \cite{borovsky_solar_2010}, a 7-minute running window was applied to subtract the background, meaning that only fluctuations on time scales smaller than 7 minutes were considered. In contrast, our use of wavelet transform allows us to select various frequency ranges, enabling the analysis of fluctuation energies across a broader range of scales. In Range 4—corresponding to time scales near 7 minutes—we do observe a slight dip in $\sigma_r$ at the zero epoch, as shown in Fig. \ref{fig:sigmas}. 
However, this dip is not as pronounced as the one reported by \citet{borovsky_solar_2010}, and it does not appear in the other frequency ranges. It is worth noting that our analysis is based on a larger  database and uses the median value instead of the mean, which helps exclude extreme values and results in smoother, more robust trends.

Compared with previous observation \citep{shi_alfvenic_2021} and simulation \citep{shi_influence_2022}, our superposed median of $\sigma_c$ shows a similar dip at the interface. 
Regarding $\sigma_r$, \citet{shi_alfvenic_2021} investigated on a velocity shear region at 0.2 AU using data from PSP, where the $\sigma_r$ displayed a distinct increase at the interface.
In that study, a Fourier transform was applied to subtract the 30-minute scale background field and extract the fluctuation energy. We note that 30 minutes is already close to the time scale of the large-scale velocity shear. 
Moreover, at 0.2 AU the compression between slow and fast wind is much weaker than at 1 AU, suggesting that the observed increase in $\sigma_r$ was likely a numerical effect induced by applying Fourier Transform within velocity shear region, rather than caused by the compression. 
In the 2.5D MHD simulation conducted by \citet{shi_influence_2022}, an increase in $\sigma_r$ is observed around the shear region even if the background velocity profile is completely removed before computing the wave energies, which is contrary to the drop of $\sigma_r$ in our superposed analysis. The discrepancy between the simulation and our current study is not fully understood yet. One possible explanation is that in the simulation the numerical dissipation rate for magnetic field is larger than the velocity due to strong phase-mixing in the shear region. 
However, a future investigation is necessary to confirm this conjecture which is beyond the scope of the current work.

\begin{figure*}[ht!]
\plotone{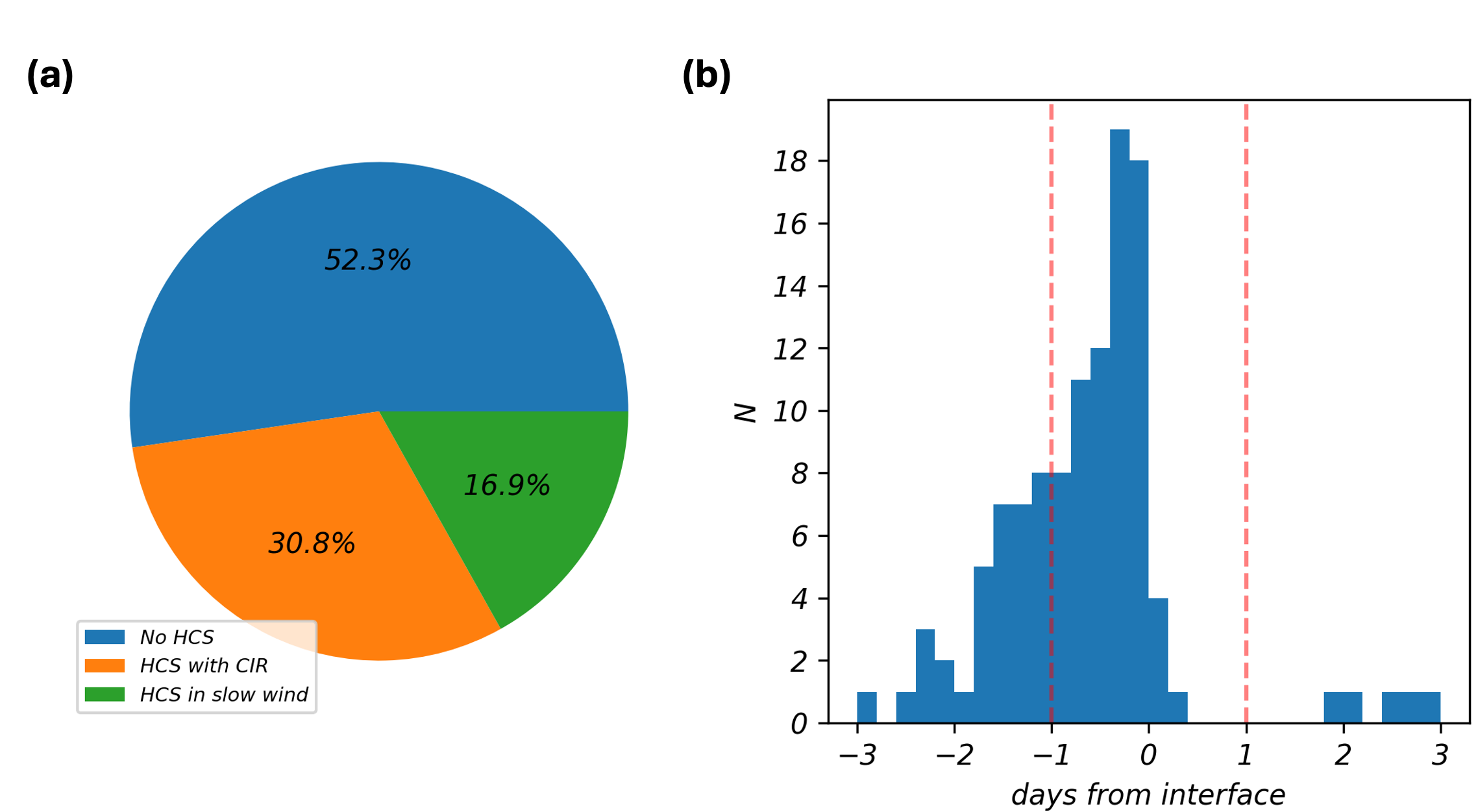}
\caption{Statistical results of the relative arrival time of HCS to the stream interface. Red dashed lines in panel (b) mark $\pm 1$ day around the stream interface, which is the threshold used to differentiate ``HCS with CIR'' and ``HCS in slow wind''.
\label{fig:HCS}}
\end{figure*}

Another factor that could contribute to the drop in $\sigma_c$ across various frequency ranges is the presence of HCS \citep{grappin_alfvenic_1991, shi_influence_2022}. 
In our analysis, we did not account for the influence of the HCS on turbulence properties, but we performed a statistical analysis to determine how many events in our dataset are associated with the HCS. The HCS is first identified based on the following criteria: (1)a significant change in the peak of the strahl electron PAD, typically shifting from $\geq150^{\circ}$ to $\leq30^{\circ}$ or vice versa, and (2) a polarity reversal of the radial component of the magnetic field \citep{behannon_fine-scale_1981, lepping_large-scale_1996, szabo_heliospheric_2020}.
To specify when the spacecraft travels through the HCS, we define the ``typical angle'' as the peak angle of PAD at each time moment. The HCS appearance time is then identified as the time moment with the highest ``typical angle'' gradient, indicating a significant shift in the PAD.
The time difference between the arrival of stream interface and arrival of HCS is listed in the supplementary material.
In our dataset, 73 CIRs are associated with HCS located near the stream interface, defined as within 24 hours of the zero epoch. 
In contrast, 35 events exhibit HCS in the slow wind, defined as occurring at least 24 hours away from the zero epoch.
Panel (a) of Fig. \ref{fig:HCS} shows the distribution of events without HCS(blue), events with HCS in slow wind(green), and events with HCS near the stream interface(orange). 
A more detailed distribution of the relative arrival time of the HCS with respect to the stream interface is shown in panel (b). Several events appear to have the HCS located two or three days after the passage of CIR. However, these are not current sheets in the fast solar wind stream. Instead, our inspection of individual cases reveal that those scattered events beyond +2 days in panel (b) are caused by the spacecraft re-entering a slow stream after passing through the CIR.
In our dataset, 46.9\% CIRs are associated with HCS considering $\pm$3 days. This rate is close to the ratio 54\% in \citep{jian_solar_2019}. In their study, they conducted a comprehensive comparison between CIRs with HCS crossings and those without.
Their results showed that CIRs associated with an HCS tend to have slightly lower speeds but exhibit higher peak values of number density, magnetic field strength, dynamic pressure, and total pressure compared to CIRs without an HCS.
However, they did not show the exact arrival time of the HCS relative to the interface.
In our results, it is noteworthy that among all CIRs accompanied by an HCS, more than half show the HCS occurring very close to the stream interface. Back to a region close to the sun, the HCS may be within the slow wind, and parallel to but not initially embedded with CIRs. However, as the fast wind propagates and the slow wind becomes compressed, the HCS gradually shifts closer to the stream interface \citep{gosling_formation_1999}. 
This may explain some cases where CIRs are accompanied by an HCS, but a substantial number of events still show the HCS located precisely at the stream interface, which indicate the slow and fast stream may originate from two regions separated by the HCS.
A clearer explanation for this intriguing correlation requires further study.

\section{Conclusions} \label{sec:conclusion}
We use WIND data and SEA to investigate the turbulence properties of over 200 CIRs at 1 AU, with the zero epoch determined through MVA. Each event features distinct shear zones, followed by a long interval of fast solar wind. The turbulence is measured by normalized cross helicity ($\sigma_c$) and normalized residual energy ($\sigma_r$), and fluctuation energies are measured by wavelet transform. The major findings are listed as below:
\begin{enumerate}
\item The spectral indices of the inertial range for both magnetic and velocity turbulence show a smooth transition within the CIR, from the steeper slow wind side to shallower fast wind side. In the dissipation range, the indices of velocity exhibit a drop before the stream interface, which could be a result of the compression.
\item Turbulence properties show different behaviors across different frequency ranges. When considering a broad frequency band, from $1.57\times10^{-5}~s^{-1}$ to $6\times10^{-3}~s^{-1}$, 
$\sigma_c$ shows a noticeable drop near the stream interface, while $\sigma_r$ shows a characteristic peak-valley-peak structure. In contrast, within a higher-frequency range that excludes low-frequency components, from $1\times10^{-4}~s^{-1}$ to $6\times10^{-3}~s^{-1}$, the drop in $\sigma_c$ remains evident, but $\sigma_r$ instead shows a smooth transition from around -0.45 to around -0.3 within the CIR. The drop in $\sigma_c$ is attributed to the enhancement of inward Alfv\'en wave energy due to compression, whereas the peak-valley-peak structure in $\sigma_r$ is primarily influenced by the large-scale velocity shear.
\item A more detailed analysis across narrower frequency bands reveals that a strong peak in $\sigma_r$ appears within the CIR in very low frequency ranges, clearly indicating that velocity shear can lead to an overestimation of velocity fluctuation energy, especially at time scales comparable to the CIR structure. In the higher-frequency ranges, $\sigma_r$ shows a general decrease across the slow wind side, CIR region, and fast wind side as the frequency decreases. Meanwhile, on the fast wind side, $\sigma_c$ displays an increase–decrease trend with decreasing frequency, reaching its maximum near the range of $1\times10^{-4}$–$6\times10^{-3}~\text{s}^{-1}$.
\item Among all CIR events in our dataset—from 2003 to 2021 at 1 AU—nearly half are accompanied by an HCS. Furthermore, of the CIRs with an associated HCS, more than half exhibit an HCS occurring within $\pm$1 day of the stream interface. A substantial number of events even show an HCS closely aligned with the interface.
\end{enumerate}

Our results help reconcile the inconsistency between previous works \citep{borovsky_solar_2010, shi_alfvenic_2021,shi_influence_2022}, and highlight the characteristics of solar wind turbulence near CIRs across various time scales, offering deeper insights into the turbulence evolution. Furthermore, we provide a valuable comparison between the distinct attributes of slow and fast solar wind streams on both sides of the CIRs. In our dataset, nearly half of the CIRs are accompanied by an HCS, which may serve as an additional factor influencing the turbulence properties. Additionally, the reasons why a large number of HCSs are perfectly aligned with the stream interface requires a comprehensive inspection and confirmation. A potential direction for future work is to check the source regions of typical HCS-related CIR events in our dataset with the help of ballistic tracing and Potential-Field Source-Surface (PFSS) model.

\section*{Acknowledgments}
This work is supported by NSF SHINE 2229566. Y.D. acknowledges NASA HTMS 80NSSC20K1275 and NASA AIAH 80NSSC25K0386. NS is supported by NASA Award 80NSSC24K0272.

%

\vspace{5mm}

\software{pyspedas\citep{grimes_space_2022}, pywavelet\citep{lee_pywavelets_2019}}




\bibliography{yuliang_CIRturbulence}{}
\bibliographystyle{aasjournal}



\end{document}